\documentclass[10pt,notitlepage,pra,twocolumn]{revtex4-1}
\usepackage{amsfonts}
\usepackage{amsmath}
\usepackage{amssymb}
\usepackage{graphicx}
\usepackage{float}
\usepackage[toc,page,header]{appendix}
\usepackage{color}
\usepackage{multirow}

\begin{document}

\title{Periodically-driven facilitated high-efficiency dissipative entanglement with Rydberg atoms}
\author{Rui Li$^{1}$, Dongmin Yu$^{1}$, Shi-Lei Su$^{2,\dagger}$, Jing Qian$^{1,*}$}
\affiliation{$^{1}$State Key Laboratory of Precision Spectroscopy, Quantum Institute for Light and Atoms, Department of Physics, School of Physics and Electronic Science, East China Normal University, Shanghai 200062, China}
\affiliation{$^{2}$ School of Physics, Zhengzhou University, Zhengzhou 450001, China}

\begin{abstract}
A time-dependent periodical field can be utilized to efficiently modify the Rabi coupling of system, exhibiting nontrivial dynamics. We propose a scheme to show that this feature can be applied for speeding up the formation of dissipative steady entanglement based on Rydberg anti-blockade mechanism in a simplified effective configuration, fundamentally stemming from a frequency match between the external-field modulation frequency and the systematic characteristic frequency. In the presence of an optimal modulation frequency that is exactly equal to the central frequency of driving field, it enables a sufficient residence time of the two-excitation Rydberg state for an irreversible spontaneous decay onto the target state, leading to an accelerated high-fidelity steady entanglement $\sim0.98$, with a shorter formation time $<400\mu$s. We show that, a global maximal fidelity benefits from a consistence of microwave-field coupling and spontaneous decay strengths, by which the scheme manifests robust insensitivities towards the imperfect initialization, the fluctuations of  vdWs interaction and modulation frequency. This simple approach to facilitate the generation of dissipative entangled two-qubit states by using periodic drivings may guide a new experimental direction in Rydberg quantum technology and quantum information.

\end{abstract}

\email{jqian1982@gmail.com}

\maketitle
\preprint{}

\section{Introduction}

Dissipative mechanism, in contrast to its natural idea that leads to the destruction of quantum effects de-coherently, can counterintuitively serve as an important resource for implementing quantum information task and controlled quantum state preparation, having been denoted into numerous studies \cite{Diehl08,Verstraete09,Cho11,Kastoryano11,Reiter12,Torre13,Reiter16,Cian19}. Experimental achievements towards dissipative production of entangled states have been implemented in trapped ions \cite{Barreiro11,Lin13}, superconducting quantum bits \cite{Shankar13,Schwartz16}, and macroscopic atomic ensembles \cite{Krauter11}.

It is remarkable that pioneer works demonstrating preparation of dissipative steady entanglement from an uppermost Rydberg level, were proposed by Saffman \cite{Carr13} and M\o lmer \cite{Rao13}, providing great potentials for applications of quantum computation and engineering by Rydberg dissipation \cite{Weimer10,Saffman16}. However typical time needed for this dissipative preparation has exceeded hundreds of milliseconds due to the use of a high-lying long-lived Rydberg level with its principle quantum number $n\sim125$ {\it e.g.} see \cite{Shao142,Su15}, 
yet difficult for the experimental measurement. So far great efforts towards ways of accelerated formation of an entangled steady state use Rydberg electromagnetically induced transparency involving a dark eigenstate, which is influenced by a fast-decaying middle state in the presence of Rydberg interactions, promising a rapid entangled-state generation \cite{Rao13}. Nevertheless this approach simultaneously suffering from the complexity of energy levels and laser fields combining with Rydberg interactions \cite{Chen18,ChenH18, Chen19}, still has not been realized in experiment.

Other alternative approaches are proposed by combining an optical cavity to trigger the entanglement formation, because the cavity decay treating as an auxiliary loss channel towards the target state, can permit a reduced stabilization time that is smaller than tens of microseconds depending on the absolute value of cavity mode coupling \cite{Shen11,Su14,Shao14,Shao16,Shao17}. More recently an intriguing improvement for a ten-times faster generation of steady entanglement 
in a cavity is suggested by using pulse modulation, achieving an exponentially enhancement for the atom-cavity coupling strength \cite{Chen19}. While such schemes based on cavity-trapped atoms require a stronger atom-cavity coupling associated with a precise control, remains uneasy for real  implementations.

In parallel, periodically driven systems are well-known in quantum physics arising a wealth of versatile quantum phenomena \cite{Denisov07,Poggi14,Yudin16,Baran19}. For example a simplest two-level atom system driven by a periodically-modulated field can significantly modify the time evolution of system \cite{Agarwal94,Noel98}, exhibiting many intriguing effects such as persistent atomic trapping in excited state \cite{Li19}, an excited two-level emitter \cite{Macovei14}, multi-photon resonance and response \cite{Manson96}, maximal population transfer \cite{Poggi14}, discrete time crystals \cite{Gambetta19,Yu19} and so on. In addition, periodically-driven dissipative (open) systems offer various prospects for new-class feature of out-of-equilibrium physics which is inaccessible for equilibrium ensembles \cite{Eisert15,Brandner16,Reimer18}, covering promising applications from topological states to experimental optomechanics \cite{Abanin15,Ge17,Shtanko18,Aranas17}. Combining periodically-driven with a Rydberg atom array is found to produce the localization of quantum many-body state \cite{Basak18}, and more recently, a new mechanism driven by an amplitude-modulated periodic field to generate dissipative steady-state entanglement in a solid-state qubit system is exploited \cite{Gramajo18}, giving more perspective for nontrivial periodically-driven Rydberg features.

Utilizing Rydberg anti-blockade mechanism, in the present work we develop a simplified scheme for the dissipative preparation of a maximal steady entangled state $|S\rangle = \frac{1}{\sqrt{2}}(|10\rangle-|01\rangle)$, driven by a periodic pump field, representing an exotic and accelerated formation. 
Remarkably, the existence of an external modulation frequency matching with the characteristic frequency of the effective two-state system, will arise a visible change to the Rabi behavior of the two-excitation Rydberg state in the laser pumping for unitary dynamics, which benefits from a fast decaying onto the target state due to existing a longer residence time. Accompanying with a suitable adjustment for other dissipative rate and microwave coupling strength, the resulting entanglement formation reveals a stepwise acceleration towards the target entangled state with a very high-fidelity $\sim0.98$, where the required convergence time can be shorter than $400\mu$s. 
Because a stepwise-improving entangled-state accumulation rate will appear within these durations, in contrast to the case of a continuous driving where the accumulation rate persists slow. Additionally we demonstrate an easy way for further acceleration by increasing the effective Rabi coupling strength in the pumping, and stress the robust insensitivities of scheme to arbitrary initial preparation, variable interaction and driving frequency under real experimental parameters.

\section{Theoretical formulation}

\subsection{Single original four-level atom}

\begin{figure}
\includegraphics[width=3.2in,height=2.0in]{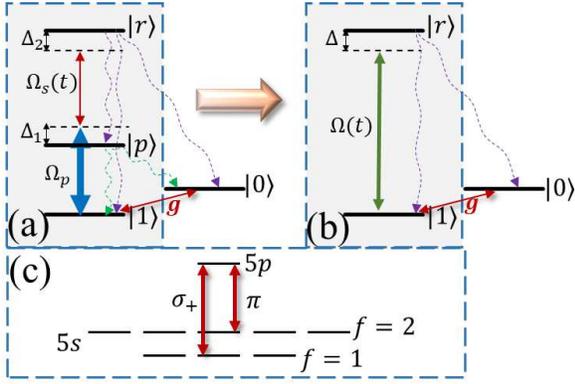}
\caption{(color online) (a-b) Schematic of an effective three-level atom deriving from its original form of a four-level configuration. See texts for detailed parameter descriptions. (c) {\it e.g.} The two hypefine ground states $|1\rangle=|5s,f=1,m_f=0\rangle$, $|0\rangle=|5s,f=2,m_f=0\rangle$ are coupled by a weakly resonant Raman coupling with respect to the middle state $|p\rangle=|5p\rangle$, or by a direct microwave coupling $g$ between $|0\rangle$ and $|1\rangle$. }
\label{modsub}
\end{figure}

To verify the validity of our effective five-level scheme as adopted in Fig.\ref{mod}(b), we first begin with an original four-level atom with ground state $|1\rangle$, middle state $|p\rangle$, Rydberg state $|r\rangle$ composing a two-photon excitation \cite{Miroshnychenko10}, where the Rabi frequencies are denoted by $\Omega_p$ and $\Omega_s(t)$ (time-dependent), the corresponding detunings are $\Delta_1$ and $\Delta_2$. State $|0\rangle$ serves as an auxiliary ground state introduced to achieve a weak microwave coupling $g$ with state $|1\rangle$. Alternatively, this could be accomplished with a Raman transitions with respect to $|p\rangle$, {\it e.g.} see Fig. \ref{modsub}(c).

Converting into the interaction picture, the Hamiltonian of single atom reads as $\mathcal{\hat{H}}_1 = \mathcal{\hat{H}}_{1d}+g/2(|1\rangle\langle0|+|0\rangle\langle1|)$ with the atom-light interaction term $\mathcal{\hat{H}}_{1d}$ taking form of
\begin{equation}
\mathcal{\hat{H}}_{1d} = \frac{\Omega_p}{2}e^{-i\Delta_1t}|p\rangle\langle1| +  \frac{\Omega_s(t)}{2}e^{-i\Delta_2t}|r\rangle\langle p| + H.c.
\label{sing}
\end{equation}
By assuming the condition of two-photon resonance $\Delta_1+\Delta_2 = 0$ by $\delta = |\Delta_{1,2}|$, if a big detuning to $|p\rangle$ is used by satisfying $\delta\gg\Omega_p,\Omega_s(t)$, it is reasonable to eliminate state $|p\rangle$ in an adiabatic way \cite{Brion07} leading to a direct coupling between $|1\rangle$ and $|r\rangle$ then the Hamiltonian $\mathcal{\hat{H}}_{1d}$ is modified to be
\begin{equation}
\mathcal{\hat{H}}_{1d}^{\prime} = \frac{\Omega_p^2}{4\delta}|1\rangle\langle1|+\frac{\Omega_s^2(t)}{4\delta}|r\rangle\langle r| + \frac{\Omega_p\Omega_s(t)}{4\delta}(|r\rangle\langle1|+|1\rangle\langle r|)
\end{equation}

After some organization the entire single-atom Hamiltonian $\mathcal{\hat{H}}_1$ can be rewritten in a reduced  form only referring to states $\{|1\rangle$, $|r\rangle$, $|0\rangle\}$, as
\begin{equation}
\mathcal{\hat{H}}_1 = -\Delta(t)|r\rangle\langle r| + (\frac{\Omega(t)}{2}|r\rangle\langle1|+\frac{g}{2}|1\rangle\langle0|+H.c.)
\label{singHam}
\end{equation}
with the two effective parameters \cite{Muller14}
\begin{equation}
\Omega(t) =\frac{ \Omega_p\Omega_s(t)}{2\delta},\Delta(t) =\frac{ \Omega_p^2-\Omega_s^2(t)}{4\delta} \nonumber
\end{equation}
represented in the Fig.\ref{modsub}(b), in which state $|1\rangle$ is straightforward excited to the highly-excited level $|r\rangle$ via a time-dependent Rabi frequency $\Omega(t)$, detuned by $\Delta$. Noting if $\Omega_p\gg|\Omega_s(t)|$ the detuning $\Delta$ approximately preserves a constant $\Omega_p^2/4\delta$ that does not depend on time $t$. The microwave coupling $g$ is too small to affect the strong atom-light interaction, safely leaving it unvaried in Eq. (\ref{singHam}). All spontaneous dissipations from $|r\rangle$, $|p\rangle$ are considered, denoted by $\gamma$, $\Gamma$, respectively. A numerical verification of the unitary dynamics by comparing effective and exact models will be left for discussion in the appendix A.

\subsection{A pair of reduced three-level atoms}

\begin{widetext}

\begin{figure}
\includegraphics[width=4.3in,height=3.0in]{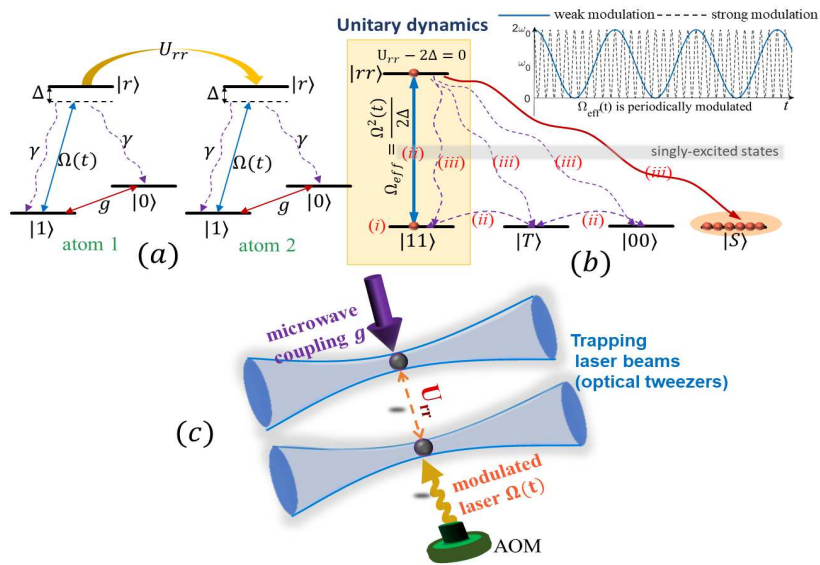}
\caption{(color online) Schematic diagram of a high-fidelity facilitated entanglement preparation. (a) A detailed two-atom energy-level structure. For each atom, two ground states $|1\rangle$ and $|0\rangle$ are coupled by a continuous microwave field $g$, and $|1\rangle$ is far-off resonantly coupled to the Rydberg state $|r\rangle$ via a periodically-modulated optical field with the Rabi frequency $\Omega(t)$. Assuming that the detuning $\Delta$ can compensate the interaction induced energy shift $U_{rr}$ by satisfying the anti-blockade condition $U_{rr} = 2\Delta$ persistently, $|11\rangle$ is directly coupled to $|rr\rangle$ via a two-photon excitation with an effective value $\Omega^2(t)/2\Delta$. (b) The effective five-level diagram from a pair of three-level atoms shows the real atom-field interactions and spontaneous decays, where the four singly-excited states (gray shadow area) are safely discarded due to $\Delta\gg\Omega(t)/\sqrt{2}$. $|S\rangle$ is a target dark entangled state, unidirectionally receiving the population from $|rr\rangle$ through spontaneous emission. The amplitude of $\Omega_{eff}(t)$ is illustrated in the inset where blue-solid and black-dashed curves stand for weak and strong frequency modulations, respectively. (c) The experimental setup (proposal, see Section \rm{V} for detailed experimental parameters). Two atoms are trapped in dipolar traps formed by tightly focused laser beams (optical tweezers) with interatomic distance $R$ close to the critical value $R_c$, leading to the strength of van der Waals-type interaction $U_{rr} = C_6/R^6$ between two individual atoms. They are also persistently driven by a microwave field $g$ as well as a periodically-modulated optical field $\Omega(t)$ realized by an acousto-optical modulator in the experimental implementation. The variation of interaction $\delta U(t)$ due to the imperfect position confinement in the foci of optical tweezers, typically $<1 \mu$m, will be discussed in section \rm{VI}(B). }
\label{mod}													
\end{figure}

\end{widetext}

We proceed to show more essences of energy structure when a pair of effective three-level atoms are involved. The relevant energy levels interested have been represented in Fig.\ref{mod} (a) composing two reduced three-level $\Lambda$ interacting atoms [Fig.\ref{modsub}(b)]. For each atom existing two ground states $|0\rangle$, $|1\rangle$ and one Rydberg state $|r\rangle$, the hyperfine states $|0\rangle$ and $|1\rangle$ suffering from dipole-forbidden are coupled by a microwave field with strength $g$, receiving the population from $|r\rangle$ through stochastic spontaneous emission decays by rate $\gamma$. Population accumulated on other hyperfine ground states rather than $|0\rangle$ and $|1\rangle$, can be re-pumped onto $|0\rangle$ or $|1\rangle$ by recycling lasers (not shown) \cite{Carr13}. Special attention is paid to the optical transition between $|1\rangle$ and $|r\rangle$ driven by a time-dependent field $\Omega(t)$, accompanied by a big detuning $\Delta$ (assuming $\Delta\gg\Omega(t)/\sqrt{2}$) for realizing  a complete suppression to the singly-excited collective states. $U_{rr}$ describes the two-atom van der Waals(vdWs)-type interaction, making it dynamically compensate the detuning of $|rr\rangle$ by $U_{rr}=2\Delta$, which can facilitate a direct resonant coupling $\Omega(t)^2/2\Delta$ between $|11\rangle$ and $|rr\rangle$ via adiabatically eliminating the middle single-excitation Rydberg states, as shown in Fig.\ref{mod}(b).

In a rotating-wave frame, the basic Hamiltonian for a pair of three-level atoms involving the single-atom Hamiltonian $\hat{\mathcal{H}}_{1}$, reads
\begin{equation}
\hat{\mathcal{H}}=\hat{\mathcal{H}}_{1}\otimes\hat{I}+\hat{I}\otimes\hat{\mathcal{H}}_{1}+\hat{\nu}
\label{totHam}
\end{equation}
where $\hat{I}$ a $3\times 3$ unit operator. $\hat{\nu}$ describes the two-atom Rydberg interaction given by $\hat{\nu}=U_{rr}\left\vert rr\right\rangle\left\langle rr\right\vert$.

Using the two-atom base vectors that contain one doubly-excited state $|rr\rangle$, four singly-excited states $|r1\rangle$, $|1r\rangle$, $|r0\rangle$, $|0r\rangle$, and four ground states $|00\rangle$, $|11\rangle$, $|T\rangle$, $|S\rangle$, the complete decay behavior (dissipation) can be expressed by the regular Lindblad operators,
\begin{eqnarray}
\hat{L}_{1}&=&\sqrt{\gamma}[\left\vert 11\right\rangle\left\langle r1\right\vert+(\left\vert T\right\rangle+\left\vert S\right\rangle)\left\langle r0\right\vert+\left\vert 1r\right\rangle\left\langle rr\right\vert] \nonumber\\
\hat{L}_{2}&=&\sqrt{\gamma}[\left\vert 11\right\rangle\left\langle 1r\right\vert+(\left\vert T\right\rangle-\left\vert S\right\rangle)\left\langle 0r\right\vert+\left\vert r1\right\rangle\left\langle rr\right\vert] \nonumber\\
\hat{L}_{3}&=&\sqrt{\gamma}[\left\vert 00\right\rangle\left\langle r0\right\vert+(\left\vert T\right\rangle-\left\vert S\right\rangle)\left\langle r1\right\vert+\left\vert 0r\right\rangle\left\langle rr\right\vert] \nonumber\\
\hat{L}_{4}&=&\sqrt{\gamma}[\left\vert 00\right\rangle\left\langle 0r\right\vert+(\left\vert T\right\rangle+\left\vert S\right\rangle)\left\langle 1r\right\vert+\left\vert r0\right\rangle\left\langle rr\right\vert] 
\end{eqnarray}
corresponding to the spontaneous decay channels of $|r\rangle\to|1\rangle$ ($\hat{L}_{1,2}$), and of $|r\rangle\to|0\rangle$ ($\hat{L}_{3,4}$), respectively. Noting that the ground states $|01\rangle$, $|10\rangle$ have been re-organized into the collective singlet states $|S\rangle=(|10\rangle-|01\rangle)/\sqrt{2}$, $|T\rangle=(|10\rangle+|01\rangle)/\sqrt{2}$ where $|S\rangle$ is a unique dark state, absolutely decoupled from other two-atom ground base vectors by $\langle S|\hat{\mathcal{H}}_1|ij\rangle\equiv0$ ($ij\in\{T,00,11\}$). 

Consequently it is reliable to prepare the maximal entangled state $|S\rangle$ through a unidirectional spontaneous loss from the doubly-excited state $|rr\rangle$ as shown by thick red arrow in Fig.\ref{mod}(b). Here the singly-excited states have been safely discarded due to a big detuning $\Delta$ with respect to $|r\rangle$, the preparation efficiency for entanglement formation mainly depends on the competition between excitation or de-excitation rates of $|11\rangle\leftrightarrow|rr\rangle$, and the unidirectional decay process ($\propto \gamma$) from $|rr\rangle$ to $|S\rangle$.

Remarkably, in the presence of a periodical modulation to the amplitude of $\Omega(t)$, the Rabi behavior between $|11\rangle$ and $|rr\rangle$ will be significantly changed, producing complex multi-frequency oscillations \cite{Noel98}. We verify that the resulting formation of target ground state $|S\rangle$ can be dramatically accelerated under an optimization for relevant frequency parameters.

\subsection{Effective five-level system}

The physical essence for robustly fast preparation based on a simplified five-level scheme [see Fig. \ref{mod}(b), note that the four singly-excited states are adiabatically discarded due to the far-off one-photon resonance],  can be understood in three steps:

(i) State initialization. Prepare two atoms in ground states {\it e.g. $|11\rangle$} (robustness of scheme to arbitrary initial states will be discussed in section \rm{VI}(A)).

(ii) Circular unitary dynamics and microwave coupling. A two-state unitary dynamics leads to exchanged population (excitation and de-excitation) by an engineered Rabi oscillation between $|11\rangle$ and $|rr\rangle$, governed by a reduced two-state Hamiltonian
\begin{equation}
\hat{\mathcal{H}}_{eff,uni}=\Omega_{eff}(t)(\left\vert 11\right\rangle+| rr\rangle)(\left\langle 11\right\vert+\left\langle rr\right\vert).
\label{unieq}
\end{equation}
where $\Omega_{eff}(t) =\Omega(t)^2/2\Delta$. Here the driving field $\Omega(t)$ is time-dependent, accompanied by a microwave coupling $g$ persistently transferring population among $|11\rangle$, $|T\rangle$ and $|00\rangle$. Notice that $|11\rangle$ and $|00\rangle$ are indirectly coupled via state $|T\rangle$. This process can be expressed by Hamiltonian $\hat{\mathcal{H}}_{eff,mw}$
\begin{equation}
\hat{\mathcal{H}}_{eff,mw}=\frac{g}{\sqrt{2}}(\left\vert 11\right\rangle\left\langle T\right\vert+\left\vert T\right\rangle\left\langle 00\right\vert)+H.c.
\end{equation}

(iii) Dissipative formation. Owing to the limited lifetime of $|rr\rangle$, in the effective five-level frame population on $|rr\rangle$ suffers from a big spontaneous loss, randomly decaying into four ground states. Once it possibly decays into the unique dark state $|S\rangle$, population will be irreversibly accumulated on it; otherwise they repeatedly experience circular steps (ii)$\leftrightarrows$(iii) by following the route  of returning to step (ii) again, until the system is finally stabilized onto state $|S\rangle$.

Shortly concluded, ideally the target entangled state $|S\rangle$ can be deterministically created for a sufficiently long time, robustly insensitive to arbitrary optical and microwave drivings. In fact this time could be endless which is impossible for real experimental measurement during a finite detection time. So achieving a fast and high-fidelity entanglement depending on a simplified and feasible protocol with Rydberg atoms remains challenge, having attracted numerous efforts in theory \cite{Chen18,ChenH18,Chen19,ChenYH17}. Presently we show a new way by implementing a periodical modulation to the amplitude of the pump field, revealing a dramatic facilitation to the process between (ii)$\leftrightarrows$(iii) circularly that breaks until the unique entangled state $|S\rangle$ is entirely populated. Our results have verified that the formation time for a high-fidelity steady entanglement can be accelerated by orders of magnitude compared with the case of no modulation.

In addition a further extension for the entanglement formation between two mesoscopic atomic ensembles can be considered by two Rydberg superatoms that allow for only single collective Rydberg excitation within the blockade radius. In fact, it is more attractive to utilize the dissipation-based method in experiment due to its deterministic implementation, and the first demonstration of dissipative entanglement generation with two ground atomic ensembles has been reported \cite{Krauter11}. As for two Rydberg superatoms, the computational basis states of each ensemble can be expressed as \cite{Beterov13,Ebert15}
\begin{eqnarray}
|G_1\rangle &=& |1_1,...,1_N\rangle \\
|G_0\rangle &=& \frac{1}{\sqrt{N}}\sum_{k=1}^{N}|1_11_2,....,0_k,...,1_N\rangle \\
|R\rangle &=& \frac{1}{\sqrt{N}}\sum_{k=1}^{N}|1_11_2,....,r_k,...,1_N\rangle
\end{eqnarray}
with two collective ground states $|G_1\rangle$, $|G_0\rangle$ and a collective singly-excited state $|R\rangle$ of the Rydberg superatom, formed by strong blockade to prohibit the multiatom excitation in the ensemble. This new basis states are convincible only when the dephasing effect of state coming from atomic motions can be negligible. Luckily as we know a typical relaxation time for the temperature of atomic ensemble at $T=50\mu K$ is about tens of millisecond \cite{Saffman05}, which is longer than the formation time required in our protocol by orders of magnitude. Hence, in the new basis spanned by $\{|G_1\rangle,|G_0\rangle,|R\rangle\}$ one can also introduce a $\Lambda$ configuration for demonstrating the Rydberg superatom \cite{Zhao18}, giving rise to the single-atom Hamiltonian $\hat{\mathcal{H}}_1$ replaced by a new form $\hat{\mathcal{H}}_{sa}$, 
\begin{equation}
\mathcal{\hat{H}}_{sa} = -\Delta(t)|R\rangle\langle R| + (\frac{\Omega(t)}{2}|R\rangle\langle G_1|+\frac{g}{2}|G_1\rangle\langle G_0|+H.c.).
\end{equation}

Here $\mathcal{\hat{H}}_{sa} $ describes a Rydberg superatom interacting with light or microwave fields, serving as a starting point for a next-step consideration of steady dissipative entanglement between two long-range $\Lambda$ Rydberg superatoms. Without loss of generality we will focus on the system with two individual atoms in the present work.

\section{Unitary dynamics}

\subsection{Frequency Modulation}

In step (ii) we consider a reduced subspace with only two states $|11\rangle$ and $|rr\rangle$ (yellow box in Fig. \ref{mod}(b)), in order to explore the frequency-modulated excitation dynamics. Intuitively when the reduced two-state system is driven via a continuous coupling, the system is a standard Rabi problem, revealing regular single-frequency oscillations (including excitation and de-excitation) of population between two states, where the oscillating frequency is exactly same to the Rabi frequency. A striking difference arises once the driving is modulated to be time-dependent dramatically modifying the Rabi behavior, leading to unexpected dynamics \cite{Glenn13}. 

First we introduce a cosinoidal driving field $\Omega(t)$, denoted as
\begin{equation}
\Omega(t) = \Omega_0\cos(\omega t)
\label{Omegaef}
\end{equation}
focusing on the regime where the external modulation frequency $\omega$ is compatible with the characteristic frequency of system.
Such a time-dependent amplitude modulation for the peak Rabi frequency $\Omega_0$ can be experimentally implemented via an acousto-optical modulator triggered via an electronic waveform generator controlling the acoustic profile, outputting shaped pulses \cite{Weiner11}.
Letting $\omega=0$ leads to $\Omega(t) = \Omega_0$, standing for the case of a continuous pump, same as considered in previous works \cite{Su15}.

A brief introduction to the derivation of effective two-state Hamiltonian (\ref{unieq}) can be understood in the subspace of $\{|11\rangle,|M\rangle,|rr\rangle\}$ with the singly-excited collective state $|M\rangle = (|1r\rangle+|r1\rangle)/\sqrt{2}$. The original two-state Hamiltonian $\hat{\mathcal{H}}_{uni}$ is 
\begin{equation}
\hat{\mathcal{H}}_{uni}=\frac{\Omega(t)}{\sqrt{2}}(\left\vert 11\right\rangle\left\langle M\right\vert e^{i\Delta t}+\left\vert M\right\rangle\left\langle rr\right\vert e^{i\Delta t}+H.c.) + \hat{\nu}
\end{equation}

Working in a rotating-wave frame with respect to the rotational transformation $\hat{U}=e^{-iU_{rr}\left\vert rr\right\rangle\left\langle rr\right\vert t}$, the above Hamiltonian can be re-expressed in a more concise form
\begin{equation}
\hat{\mathcal{H}}_{eff,uni}^{\prime}=\frac{\Omega(t)^2}{2\Delta}(\left\vert 11\right\rangle\left\langle rr\right\vert+\left\vert rr\right\rangle\left\langle 11\right\vert)-\frac{\Omega(t)^2}{\Delta}\left\vert M\right\rangle\left\langle M\right\vert,
\label{uneq2}
\end{equation}
where $U_{rr}=2\Delta$ and $\Delta\gg{\Omega(t)}/\sqrt{2}$ are assumed. Comparing (\ref{unieq}) and (\ref{uneq2}), we ignore the second term $-\frac{\Omega(t)^2}{\Delta}\left\vert M\right\rangle\left\langle M\right\vert$ owing to its decoupling effect of state $|M\rangle$. The effective coupling strength $\frac{\Omega(t)^2}{2\Delta}$ between $|11\rangle$ and $|rr\rangle$ can be expressed as
\begin{equation}
\Omega_{eff}(t)=\frac{\Omega_0^2}{4\Delta}+\frac{\Omega_0^2}{4\Delta}\cos(2\omega t)
\label{effecto}
\end{equation}
which includes base frequency component $\omega_0 = \frac{\Omega_0^2}{4\Delta}$ and the real-time modulated component $\omega_0\cos(2\omega t)$, giving to the peak amplitude belonging to $[0,2\omega_0]$ as shown in the inset of Fig.\ref{mod}(b). If $\omega\gg\omega_0$, the modulation term adds a fast high-frequency oscillation to the base frequency $\omega_0$ whose effect can be averagely canceled within a sufficient time, {\it i.e.} $\int_0^{\infty}\omega_0\cos(2\omega t)dt=(\omega_0/2\omega)\sin(2\omega t)\vert_0^{\infty}\to 0$, arising a dominant base frequency $\omega_0$ only; otherwise the non-negligible modulated component will induce several modulated-frequency sidebands with separation $\omega$ to the central frequency $\omega_0$, revealing complex multi-frequency dynamical behavior.

\subsection{Frequency Spectrum analysis}

 \begin{figure}
\includegraphics[width=3.0in,height=4.0in]{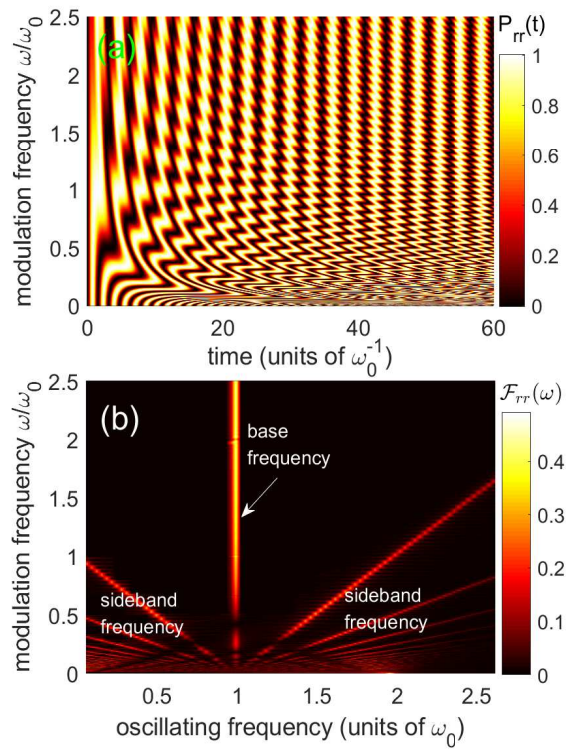}
\caption{(color online) The global output for (a) $P_{rr}(t)$ and (b) $\mathcal{F}_{rr}(\omega)$ versus the time $t\in(0,60)$ and the relative modulation frequency $\omega/\omega_0\in(0,2.5)$. Parameters are $\Omega_0 =57.6$, $U_{rr} = 1658.88$, $\Delta =829.44$ and $\omega_0$($\omega_0^{-1}$) is the frequency(time) unit. Dissipation is ignored in calculating the unitary dynamics.}
\label{spec}
\end{figure}

By numerically solving the master equation $\dot{\rho}_{uni} = i[\hat{\rho}_{uni},\hat{\mathcal{H}}_{eff,uni}]$, we detect the observable quantity $P_{rr}(t)$
\begin{equation}
P_{rr}(t) = \langle rr|\hat{\rho}_{uni}(t)|rr\rangle
\end{equation} 
for the time-dependent population probability on state $|rr\rangle$ where $\hat{\rho}_{uni}(t)$ is the density matrix of two-state subspace, as globally plotted in Figure \ref{spec}(a) versus time $t$ and the relative modulation frequency $\omega/\omega_0$. Besides Fig.\ref{spec}(b) describes a global view for the frequency spectrum of dynamics by implementing a Fourier transform $\mathcal{F}_{rr}(\omega) = \int_{-\infty}^{+\infty}{P_{rr}(t)}e^{-i\omega t}dt$. 
From Eq.(\ref{effecto}) and Fig.\ref{spec}(b) it is obvious that the presence of modulation $\omega$ can add frequency sidebands to the base frequency $\omega_0$ which is also the characteristic frequency of system, giving rise to the multiple frequencies $\omega_n = \omega_0\pm n\omega$ with $n =0,1,2 ...$, symmetrically located with respect to $\omega_0$ \cite{Ficek01}. As $\omega$ increases the dynamics become regular with one dominant base frequency $\omega_0$ because a high-modulation-frequency $\omega$ will cause an average cancellation to the accumulated quantity by the frequency modulation $|(\omega_0/2\omega)\sin(2\omega t)_{t\to\infty}|\to0$ as $\omega\gg\omega_0$. The inset of Fig.\ref{mod}(b) comparably shows the Rabi oscillation behavior of $\Omega_{eff}$ under weak (blue-solid) and strong (black-dashed) modulation frequencies.

\begin{figure}
\includegraphics[width=3.0in,height=3.3in]{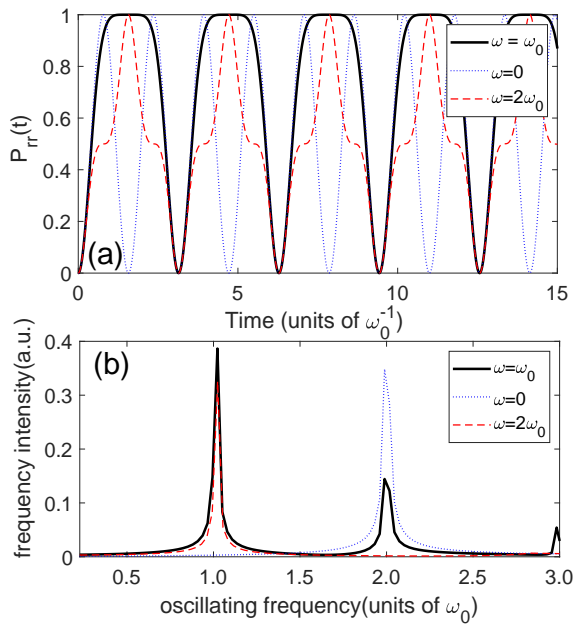}
\caption{(color online) (a-b) The unitary dynamics $P_{rr}(t)$ and frequency spectrum $\mathcal{F}_{rr}(\omega)$ are comparably presented for $\omega = 0$ (blue-dotted), $\omega =\omega_0$ (black-solid), $\omega =2\omega_0$ (red-dashed), within a short time period of $\omega_0t\in(0,15)$. Other parameters are same as adopted in Fig.\ref{spec}. }
\label{exam}
\end{figure}

As far as we know when the modulation is exactly an integral multiple of characteristic frequency $\omega_0$, {\it i.e.} $\omega = n\omega_0$, the system exists a dramatic frequency match, revealing unexpected behavior. To show this, we select $n = 0,1,2$ and represent the corresponding unitary dynamics and frequency spectrum in Fig.\ref{exam}(a-b). Except for $\omega=0$ that it reveals a complete single-frequency Rabi oscillation with frequency $2\omega_0$ ($\Omega_{eff}(t)\equiv2\omega_0$), the external frequency modulation $\omega$[$\neq 0$] will arise sideband frequencies aside from the base frequency $\omega_0$, with a tunable separation $\omega$. The resulting unitary dynamics changes significantly by meeting different frequency matching conditions.

Comparing to the case of no modulation, it is observable that for $\omega = \omega_0$ (black-solid) the population dynamics $P_{rr}(t)$ benefits from a sufficiently longer residence sustaining on the uppermost two-excitation state $|rr\rangle$, promising an efficient population dissipation onto the target state through spontaneous emission as long as the decay rate $\gamma$ is suitable. Hence the frequency match offers an essential reason for the accelerated dissipative formation in step (iii). 
Furthermore as $\omega$ increases, the system again tends to a single-frequency oscillation with the oscillating frequency $\omega_0$, revealing regular modulated Rabi behavior. For example, if $\omega=2\omega_0$ (red-dashed) $P_{rr}(t)$ suffers from a faster excitation and de-excitation processes without any stagnation on the upper state, leading to the population exchange in step (ii) between $|11\rangle\leftrightarrow |rr\rangle$ repeatedly. The resulting dissipative entanglement preparation is quite inefficient.

Based on the above analysis, we will adopt this optimal resonant modulation frequency $\omega = \omega_0$ for studying the accelerated entanglement preparation.

%An amplified picture plotted in Fig.\ref{unieq}(b) for the time range $t\in[0,50]\mu$s qualitatively gives a clear comparison for the cases (a1-a3). We observe that for $\omega=\omega_0$ the atoms are trapped in excited and ground states with same time intervals, revealing a square-wave behavior \cite{Agarwal94}. Note that here the time interval for ground-state population is wasted since no population will be accumulated on target state. Oppositely, results for the case $\omega=\omega_0/4$ show that it benefits from a fast excitation $|11\rangle\to|rr\rangle$ and simultaneously a slower de-excitation $|rr\rangle\to|11\rangle$ (denoted by green box in (a3)) in the second half of period, comparing with other cases, which can be expected for an accelerated formation of entanglement state $|S\rangle$. Although the first half of period may be still unsuited for entanglement generation since a reversed excitation and de-excitation rate is naturally predicted. We will give a further modification for this special pulse in section ?.

\section{Facilitated entanglement formation}

\begin{figure}
\includegraphics[width=3.2in,height=3.6in]{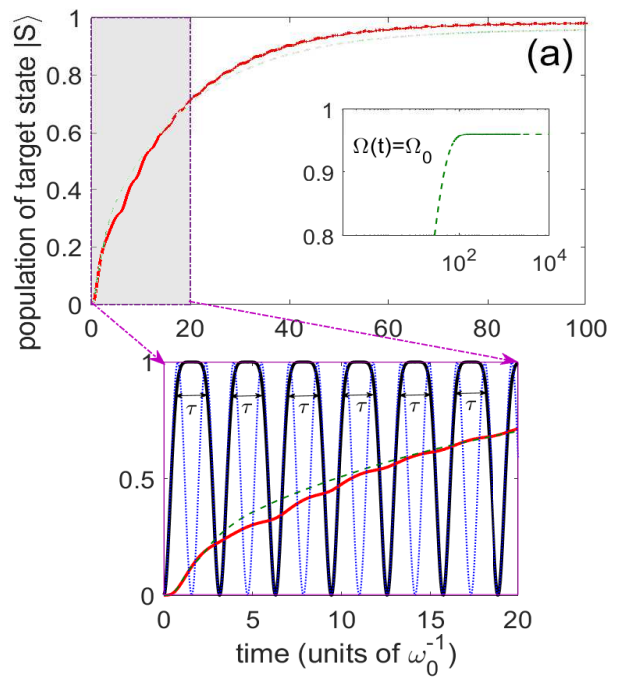}
\caption{(color online) Population of target entangled state $P_S(t)$ versus time $t$ under the modulation frequency $\omega=\omega_0$ (red-solid) and $\omega=0$ (green-dashed). A detailed plot of the relationship of the accumulated population in every duration $\tau$ on $|rr\rangle$ and unitary dynamics $P_{rr}(t)$ are comparably represented for time $t\in[0,20]$ in the inset below.
Inset inside shows an extensive plot by enlarging the time range to $10^4$ where the final population for the case of no modulation $\Omega(t)=\Omega_0$ is kept to be a saturation $\sim 0.96$. Except the optimal parameters as discussed in section \rm{III}, here we choose $\gamma = 0.4$, $g = 0.85$ and set $P_{11}(t=0)=1.0$ at the initial time. }
\label{accengle}
\end{figure}

The unitary dynamic evolution of the reduced two-state model can be guided to solve the complete master equation $\dot{\rho}=-i[\hat{\mathcal{H}},\hat{\rho}]+\mathcal{L}[{\hat{\rho}}]$ for the entire system of a pair of three-level atoms, with the basic Hamiltonian $\hat{\mathcal{H}}$ [see Eq.(\ref{totHam})] as well as the Lindblad dissipation operators, described by
\begin{equation}
\mathcal{L}[{\hat{\rho}}]=\sum_{i=1}^{4}[\hat{L}_{i}\hat{\rho}\hat{L}_{i}^{\dagger}-\frac{1}{2}(\hat{L}^{\dagger}_{i}\hat{L}_{i}\rho+\rho\hat{L}^{\dagger}_{i}\hat{L}_{i})]
\end{equation}
presenting the four spontaneous decay channels as shown in Fig.\ref{mod}(a). According to the above discussion it is confirmed that $|S\rangle$ is an absolute unique steady state, arising the system ideally staying on that state as long as the evolution time is sufficient. However, due to the competition between unitary dynamics and spontaneous dissipative process, the formation time can be endless which is far beyond the detection time in a real implementation. Remember the typical time for entanglement formation basing on similar schemes is more than tens of milliseconds by using continuous driving with higher Rydberg levels $n>100$ \cite{Chen17}. Here, with the help of a periodic amplitude modulation to the pump laser it is observed that a clear entanglement facilitation can be verified, accompanied by a reduced detecting time $\sim100/\omega_0$.

\begin{figure}
\includegraphics[width=3.0in,height=2.0in]{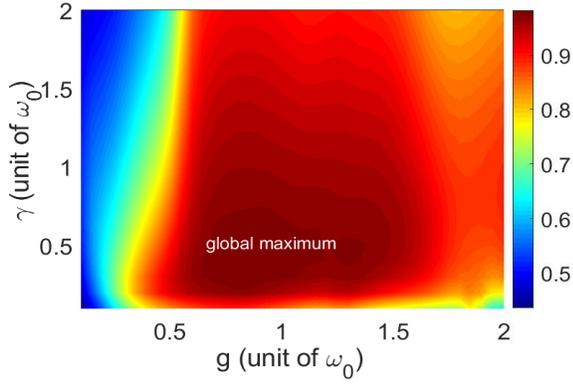}
\caption{(color online) The fidelity of final target entangled state $F_S$[$=P_S(t=100)$] versus the simultaneous adjustments for microwave coupling $g$ and spontaneous decay $\gamma$. A global maximum of $F_S$ is denoted in the plot. Here $\omega_0$($\omega_0^{-1}$) is the frequency(time) unit.}
\label{gmg}
\end{figure}

Figure \ref{accengle} exhibits the real-time dynamics for population on target state $|S\rangle$, defined by $P_S(t)=\langle S|\hat{\rho}(t)|S\rangle$, with the modulation frequency $\omega=\omega_0$ (red-solid) or without modulation $\omega=0$ (green-dashed). Clearly without periodical modulation, $P_S(t)$ represents a smoothly increasing curve but saturating towards 0.96 for $t\to\infty$, see the inset inside for an extensive range within $t\in[0,10^4]/\omega_0$. However with $\omega=\omega_0$, $P_S(t)$ exhibits a fast stepwise increase, even catching up with the case of $\omega=0$ at $t=14.3/\omega_0$, finally being stabilized to be as high as 0.98 which benefits from a significantly shortened time. Because for $\omega=0$ it is impossible to reach 0.98 during the finite detection time. The reason accounting for this facilitation can be understood by finding longer durations $\tau$ in every oscillation period on state $|rr\rangle$ that is able to provide enough time of dissipation into the target state. As indicated clearly in the inset below, during each duration time $\tau$ ($P_{rr}(\tau)=1.0$), $P_S(t)$ exhibits a dramatic enhancement especially at the initial time when the sufficient population can be exchanged between $|11\rangle$ and $|rr\rangle$.

Also we stress the importance of competition among the unitary dynamics, spontaneous dissipation and microwave coupling that leads to an accelerated entanglement formation. Therefore we globally change the rates of dissipation $\gamma$ and microwave coupling $g$, in order to see the essential importance of frequency match to the fidelity of entanglement, characterized by the final fidelity $F_S=\langle S|\hat{\rho}(t=100/\omega_0)|S\rangle$. In Fig.\ref{gmg}, by varying $\gamma$ and $g$ within a same range $(0.1,2.0)\omega_0$ there exists a global maximum region with $F_S\approx0.98$ persistently, where the values of $\gamma$ and $g$ have a best matching. In contrast, beyond that range $F_S$ reveals a considerable reduction, especially for a small $g$ value arising a significantly poor transfer rate among $|11\rangle$, $|T\rangle$, $|00\rangle$ that can not catch up with the (de-)excitation and dissipation rates, leading to a very low entanglement production $F_S<0.5$. In other words, the realization of an accelerated entanglement formation needs an optimal modulation frequency $\omega$ implemented by an external pump laser, together with a perfect frequency consistence between the dissipative rate $\gamma$ and microwave coupling strength $g$.

\section{Experimental feasibility for further facilitation}

\begin{figure}
\includegraphics[width=3.3in,height=2.4in]{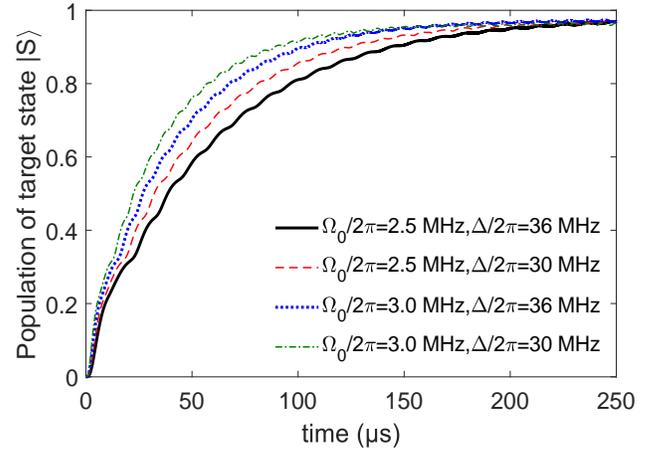}
\caption{(color online)  Time-dependent population of the target state $P_S(t)$ versus $t$($\mu$s) for different $\Omega_0$ and $\Delta$ values. Note that the conditions of $U_{rr}=2\Delta$ and $\omega=\omega_0$ are kept. }
\label{facilit}
\end{figure}

In experiment the configuration like Fig.\ref{mod}(a) can be implemented in two rubidium atoms where the hyperfine energy levels are $|1\rangle=|5s_{1/2},f=1,m=0\rangle$, $|0\rangle=|5s_{1/2},f=2,m=0\rangle$ that can be excited to the Rydberg state $|r\rangle=|100s\rangle$ via a two-photon transition mediated by {\it e.g.} $|5p_{1/2}\rangle$. A practical calculation involving $|5p_{1/2}\rangle$ would be considered in the Appendix A. Here the effective two-photon Rabi frequency is achieved to be a reasonable value $\Omega_0 = 2\pi\times2.5$MHz \cite{Shao17}, with which the ground state $|1\rangle$ is coupled to the $|r\rangle$ by a big detuning $\Delta = 2\pi\times36$MHz, giving rise to the characteristic frequency of system $\omega_0 =\Omega_0^2/4\Delta= 0.273$MHz. In the design two atoms can be trapped separately with an interatomic distance $R = 9.5\mu$m for realizing a considerable vdWs interaction $U_{rr} = 2\pi\times 72$MHz ($C_6/2\pi =5.3\times10^{13} $s$^{-1}\mu$m$^6$), driven by a periodically time-dependent laser beam $\Omega(t)$ with amplitude $\Omega_0$ and modulation frequency $\omega = \omega_0 = 0.273$MHz. Under an optimization for other frequency parameters we have $\gamma = 2\pi\times 17$kHz \cite{Beterov09} and $g =2\pi\times36.9 $kHz that enable a well coincidence with the unitary dynamics, the final fidelity for entangled state formation can attain as high as $F_S=0.981$ within a shortened operation time $T_S = 366.3\mu$s. From Fig. \ref{gmg} these values $\gamma$ and $g$ can also be tunable in a global maximum region, {\it e.g.} $\gamma=2\pi\times10$kHz and $g=2\pi\times 34.8$kHz, leading to $F_S=0.977$ within a same time $T_S$.

In addition, noting that the optimal dynamics of $P_S(t)$ as displayed in Fig.\ref{accengle}(red solid) reveals a stepwise-increasing behavior, dominantly decided by the increasing rate during the residence time $\tau$ on $|rr\rangle$ in the case of unitary dynamics. As a result an intuitive way for further facilitation to entanglement is increasing the effective coupling strength $\omega_0$ between $|11\rangle$ and $|rr\rangle$, accelerating the accumulated rate of population on the target state during each residence. Figure \ref{facilit} exhibits that, if $\Omega_0$ increases or $\Delta$ decreases, leading to a stronger coupling value $\omega_0$ for the excitation and de-excitation transitions during the laser pumping process, $P_S(t)$ can achieve a further accelerated growth, quickly saturating to a high-fidelity value $\sim 0.98$ within a shorter time. It is confirmed that the case denoted by a green-dashed curve under $(\Omega_0,\Delta)=2\pi\times(3.0,30)$MHz shows a fastest speed to saturation. That fact provides an easy facilitation optimization in this scheme when other relevant frequency parameters are already suitable.

%As verified by Fig. \ref{facilit} (a-b) we respectively change $\Omega_0$ and $\Delta$ in order to see the relationship between the formation time $T_S$ as well as the oscillation frequency for unitary dynamics. Fig.\ref{facilit}(a) shows $T_S$ versus the variation $\delta\Omega_0$ and it is found $T_S$ can be further facilitated below 484.9$\mu$s via a positive value {\it i.e.} $\delta\Omega_0/\Omega_0\in[0.02,0.16]$ since the oscillation frequency is $\omega_0\propto\Omega_0$. In an opposite case, if $\delta\Delta/\Delta\in[-0.2,-0.04]$ there also exists a small range for the formation facilitation because the optimal oscillation frequency $0.25\omega_0$ is inversely proportional to the detuning $\Delta$. It is remarkable that the facilitation mechanism based on the adjustment for oscillation frequency by tuning $\Omega_0$ or $\Delta$ is quite limited here, where the maximal facilitation of formation time $T_S$ is predicted to be $\sim 450\mu$s.

\section{Robust insensitivity of scheme}

\subsection{Imperfect initialization}

\begin{figure}
\includegraphics[width=3.4in,height=2.5in]{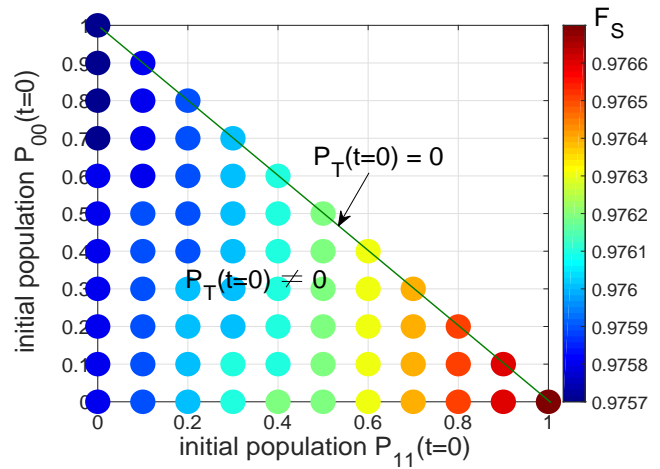}
\caption{(color online) The final fidelity $\bar{F}_S$ for a duration of $t\in[T_S-100,T_S]\mu$s, versus the variation of initial population probability on ground states $|00\rangle$, $|T\rangle$ and $|11\rangle$, defined by $P_{00}(t=0)$, $P_{T}(t=0)$ and $P_{11}(t=0)$, satisfying the conserved normalization $P_{11}(0)+P_T(0)+P_{00}(0)=1.0$. The target state $|S\rangle$ is initially unoccupied. }
\label{ini}
\end{figure}

For exploring the relation of initial population distribution and the final fidelity of the target state we show the final fidelity under different initial population $P_{00}(0)$, $P_{T}(0)$ and $P_{11}(0)$, meeting the normalized condition $P_{11}(0)+P_T(0)+P_{00}(0)=1.0$. It needs to point out that in plotting Figure \ref{ini} we have calculated an average $\bar{F}_S$ value covering a duration of $100\mu$s for overcoming the slight population fluctuations in its dynamics towards the saturation, which comes from a modified Rabi oscillation as in Fig.\ref{accengle}.

For atom initially occupying a determined state $|11\rangle$ {\it i.e.} $P_{11}(0)$=1.0 the unitary dynamics in step (ii) between $|11\rangle$ and $|rr\rangle$ is straightforward, leading to a maximal average fidelity $\bar{F}_S\sim0.9767$, as indicated by a dark red dot in Fig.\ref{ini}. However, in an opposite case if atoms are entirely prepared in $|00\rangle$ that promises a fully indirect transfer to $|11\rangle$ mediated by $|T\rangle$, a very small decrease by an amplitude of $\Delta F_S\sim$0.001 is observed for the average fidelity, strongly proving the robustness of fidelity insensitivity against this imperfect initialization. Otherwise when the initial population is prepared imperfectly in a superposition state of $|00\rangle$, $|T\rangle$, $|11\rangle$ except the target state $|S\rangle$, a continuous coupling by the microwave field $g$ among them would suffer from a persistent population transfer towards $|11\rangle$, causing a gradual varying of average fidelity between the former two cases. Intuitively {\it e.g.}, for a given $P_{00}$ value, $F_S$ continuously grows with the increase of $P_{11}$ which is directly connected to $|rr\rangle$. 
Above results again stress the flexibility and insensitivity of our scheme with respect to the system imperfect initialization, promising a feasible way for the generation of steady entanglement in current experimental environment.

\subsection{Variable vdWs interactions}

Consider in a real implementation, the pair of atoms can be individually trapped in optical tweezers separated by a tunable distance $R$, 
enabled by changing the incidence angle of the formation beams. That achieves a large vdWs interaction between the doubly-excited Rydberg state $|rr\rangle$, well compensated by the detuning $\Delta$ {\it i.e.} $2\Delta = U_{rr} = C_6/R^6$ (antiblockade). Current technique can control the distance $R$ between the two atoms with $\mu$m-scale accuracy, however it is still possible to slightly vary the positions of the individual atoms under 1 $\mu$m during the entanglement formation of a few hundreds of $\mu$s, actually making the antiblockade condition unpreserved. 

\begin{figure}
\includegraphics[width=3.4in,height=2.1in]{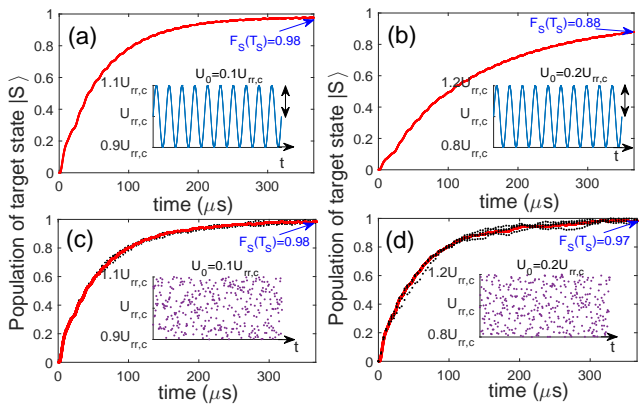}
\caption{(color online) The real time dependence of population $P_S(t)$ (red solid) on the target state $|S\rangle$ with variable vdWs interactions $U_{rr}(t)$ realized by a slight change $\delta U(t)$ around its critical value $U_{rr,c}$. In case (i) $\delta U(t)$ shows a sinusoidal function with amplitude (a) $U_0 = 0.1U_{rr,c}$ and (b) $U_0 = 0.2U_{rr,c}$. Turning to case (ii), $\delta U(t)$ is obtained randomly from $[-U_0,U_0]$ with (c) $U_0 = 0.1U_{rr,c}$ and (d) $U_0 = 0.2U_{rr,c}$.  }
\label{deltaU}
\end{figure}

To simulate the realistic dynamics under a variable interaction coming from the slight change of two-atom distance, we add a time-dependent fluctuation $\delta U(t)$ to the critical value $U_{rr,c}$ by meeting $U_{rr}(t) =U_{rr,c}+ \delta U(t)$. Here the exact antiblockade relation becomes $ U_{rr,c}=2\Delta $. 
For comparison the variation $\delta U(t)$ is assumed to take form of 

(i) a regular sinusoidal function 
\begin{equation}
\delta U(t) = U_0\sin(\omega_0^{\prime} t)
\label{sinq}
\end{equation}
with the frequency $\omega_0^{\prime} = 200\omega_0$;

(ii) a stochastic function $\delta U(t)$ generated from 
\begin{equation}
\delta U(t)\in[-U_0,U_0]
\end{equation}
under small fluctuation amplitudes Fig.\ref{deltaU}(a,c) $U_0 = 0.1U_{rr,c}$ and Fig.\ref{deltaU}(b,d) $U_0 = 0.2U_{rr,c}$, corresponding to the slight change of relative positions around $0.32\mu$m and $0.64\mu$m, respectively. Based on the assumptions we calculate the population dynamics $F_S(t)$ with time (units of $\mu$s) on the target state $|S\rangle$, with respect to the time-dependent variation of interaction $\delta U(t)$ [see insets] for different cases (i) and (ii), as represented in Fig. \ref{deltaU}(a-d).

By carrying out a regular sinusoidal modification to the two-atom interaction as in case (i), the population of target state reveals a clear fall with the increase of modified amplitude $U_0$, which is verified by comparing Fig.\ref{deltaU}(a) and (b) that the final fidelity $F_S(T_S)$ is only 0.88 when $U_0$ is increased to 0.2$U_{rr,c}$. However, the real fluctuation of two-atom distance in experiment is unpredictable leading to a better consideration of stochastic change of vdWs interaction between the two atoms, as represented in (c-d) where random fluctuations $\delta U(t)\in[-U_0,U_0]$ are adopted. In the calculations we individually generate five sets of random data with (c) $U_0=0.1U_{rr,c}$ and (d) $U_0 = 0.2U_{rr,c}$, and the final target population labeled by red curves is achieved by a numerical fitting of all data (black dots). In fact it is remarkable that the final fidelity of entangled state $|S\rangle$ has a surprising enhancement when the fluctuation of interactions is random in experiment, confirmed by (d) where $F_S(T_S)$ attains 0.97 [note that $F_S(F_S)=0.88$ in (b)] by using $U_0=0.2U_{rr,c}$, promising a perfect insensitivity of robust entanglement generation to a slight change of two-atom distance
(variable interactions) in a real implementation of our scheme.

\subsection{Deviation of periodic driving frequency}

\begin{figure}
\includegraphics[width=3.1in,height=2.3in]{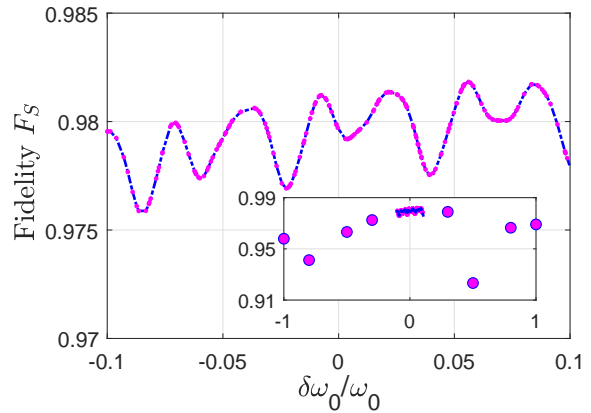}
\caption{(color online) The final fidelity $F_S$ at $T_S=366.3\mu$s, versus a small variation of perturbation $\delta\omega_0$ to the periodic driving frequency $\omega_0$, denoted by $\delta\omega_0/\omega_0\in[-0.1,0.1]$, {\it i.e.} the driving frequency changes from $0.9\omega_0$ to $1.1\omega_0$. Inset stands for an extensive range by modifying the range to $\delta\omega_0/\omega_0\in[-1.0,1.0]$.}
\label{perod}
\end{figure}

To show the robustness of scheme we consider it works under small deviations $\delta\omega_0$ of the periodic driving from its base modification frequency $\omega_0$, and verify a powerful insensitivity towards the variation of driving frequency as presented in Fig. \ref{perod}. In the calculation by adding a small deviation $\pm0.1\omega_0$ achieving the change of driving frequency from $0.9\omega_0$ to $1.1\omega_0$, the final fidelity $F_S$ reveals a slight oscillation with the peak-peak oscillating amplitude smaller than 0.005 which offers a flexible way to determine the value of modification frequency in a realistic implementation. A bigger deviation of driving frequency extending to the range of $\delta\omega_0/\omega_0\in[-1.0,1.0]$ will cause a deep fall of the output fidelity accompanied by a stronger oscillation with amplitude one order of magnitude larger, see the inset of Fig. \ref{perod}, again confirming the importance of using suitable driving frequencies that can facilitate not only the entanglement production efficiency but also a fast speed towards the final steady state.

%\section{Experimental implementation}

%For $^{87}$Rb atom at room temperature, the Rydberg level is $|r\rangle = |100s\rangle$

\section{Conclusion}

Originating from a pair of practical two four-level $\Lambda$-type Rydberg atoms, we show that when a periodically-modulated pump laser is used one can achieve an accelerated formation of dissipative entangled steady state, arising an unprecedented facilitation mechanism that has never been considered in previous similar schemes. The merit of our scheme lies on the chosen of an external modulation frequency that well agrees with the characteristic frequency of system, resulting in a dramatic modulation to the behavior of unitary dynamics in the optical pumping process. The modified Rabi oscillation behavior benefits from a longer residence time on the two-excitation Rydberg state, promising a fast decaying onto the target entangled state when the spontaneous emission rate and the microwave transferring rate are both tuned to be consistent at the same time. Under parameter optimization we successfully raise the fidelity to $\sim0.98$ with a formation time shorter than 400$\mu$s. Comparably if no modulation is carried out it is found that the fidelity persists $\sim0.96$ for a sufficient time of tens of milliseconds, un-enabling the realization of a higher entanglement within a limited detection time. Additionally we propose the way for further accelerating the convergence time by enhancing the effective coupling strength in the optical pump, and put forward to detailed discussions of the robust insensitivity of fidelity against the imperfect preparation of initial states, the variable vdWs interactions as well as the deviation of modulated frequency in a real implementation.

For realizing a fast and high-fidelity dissipative steady entanglement, our new proposal offers one-step closer to this goal, simultaneously overcoming the obstacles from a complex energy-level structure (EIT approaches) or a long formation time (traditional Rydberg anti-blockade approaches) in a number of previous works, which may provide new prospectives for experimentalists to create maximal and deterministic steady entangled state via dissipation in interacting Rydberg systems.

\acknowledgements

This work was supported by the NSFC under Grants No. 11474094, No. 11104076, No. 11804308, by the China Postdoctoral Science Foundation No. 2018T110735, by the Science and Technology Commission of Shanghai Municipality under Grants No. 18ZR1412800 and by the Academic Competence Funds for the outstanding doctoral students under YBNLTS2019-023.

\appendix

\section{Unitary dynamics of four-level atoms}

\begin{figure}
\includegraphics[width=3.4in,height=2.9in]{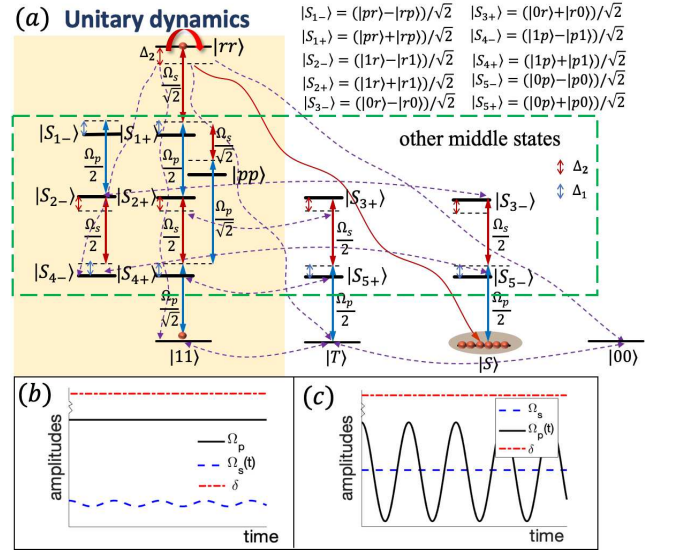}
\caption{(color online) The entire energy-level diagram of a pair of original four-level atoms. The unitary dynamics between $|11\rangle$ and $|rr\rangle$ as shown in the yellow box is mediated by two laser fields $\Omega_{p,s}$ and detunings $\Delta_{1,2}$. An effective five-level model that consists of states $\{|11\rangle,|T\rangle,|S\rangle,|00\rangle,|rr\rangle\}$ is finally obtained in the maintext, by discarding all middle states for the off-resonant detunings $|\Delta_{1,2}|\gg\Omega_p,\Omega_s$. (b-c) Two different designs of laser fields $\Omega_{p,s}$ and detuning $\delta = |\Delta_{1,2}|$, coinciding with the parameters $\Omega(t)$, $\Delta(t)$ in the effective two three-level atom system.}
\label{mapp}
\end{figure}

In order to verify the feasibility of our simpler scheme in the text, we carry out a comparable calculation for the unitary dynamics between $|11\rangle$ and $|rr\rangle$ based on the original level structures [see Fig.\ref{mapp}(a)], adopted from two rubidium atoms with states $|1\rangle =|5s_{1/2},f=1,m=0\rangle$, $|0\rangle=|5s_{1/2},f=2,m=0\rangle$, $|p\rangle =|5p_{1/2}\rangle $, $|r\rangle=|100s\rangle$. Rabi frequencies $\Omega_p$ and $\Omega_s$ are used for characterizing the optically couplings of $|1\rangle\leftrightarrows |p\rangle$ and $|p\rangle\leftrightarrows |r\rangle$ transitions. Here, the microwave coupling as well as all decays labeled by purple dashed curves in Fig.\ref{mapp}(a), are ignored.
Remember in the frame of effective system one has introduced two important parameters $\Omega(t) = \Omega_p\Omega_s/2\delta = \Omega_0\cos(\omega_0t)$ and $\Delta(t) = (\Omega_p^2-\Omega_s^2)/4\delta\approx $ const [compensated by $U_{rr}=2\Delta$] by which a high-fidelity entangled state $|S\rangle$ can be finally attained due to dissipation. Returning to the frame of an original atom-field interaction it requires the condition  of $\delta = |\Delta_{1,2}|\gg\Omega_{p,s}$ and $\Delta_1=-\Delta_2 = \delta$. For comparison here we adopt two sets of parameters $\Omega_p$, $\Omega_s$, $\delta$, as represented in Fig.\ref{mapp}(b-c) which are (b)
\begin{eqnarray}
\Omega_s(t) &=& \Omega_{s0}\cos(\omega_0t)  \nonumber\\
\Omega_p  &=& \Omega_{p0} 
\end{eqnarray}
and (c)
\begin{eqnarray}
\Omega_p(t) &=& \Omega_{p0}\cos(\omega_0t)  \nonumber\\
\Omega_s  &=& \Omega_{s0} \end{eqnarray}
with $\Omega_{p0}/2\pi = 2.88$GHz, $\Omega_{s0}/2\pi = 100$MHz, $\delta/2\pi = 57.7$GHz, $\omega_0 = 0.273$MHz. Note that $U_{rr} = 2\Delta$ is set to be preserved in the calculation no matter $\Delta$ is time-dependent or not. Cases (b) and (c) only differ by a periodic modulation term $\cos(\omega_0 t)$ whether it is implemented on the weak field $\Omega_s$ or the strong field $\Omega_p$. 

\begin{figure}
\includegraphics[width=3.1in,height=2.8in]{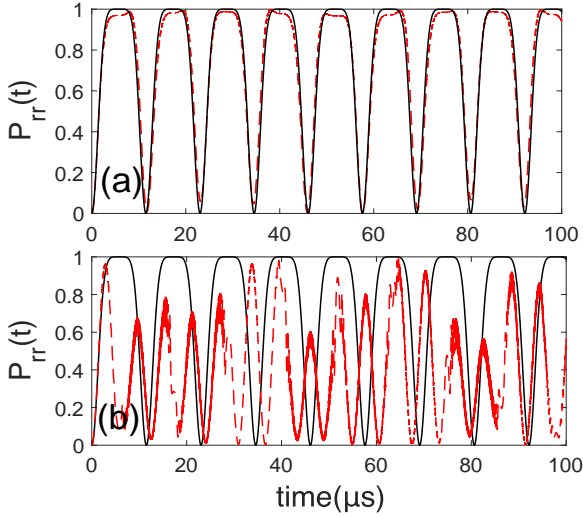}
\caption{(color online) The unitary dynamics of $P_{rr}(t)$ (red dashed) in an original pair of four-level atoms, versus time (units of $\mu$s) under (a) a periodically-modulated field $\Omega_s(t)$ and (b) a periodically-modulated field $\Omega_p(t)$. Parameters are described in the text. The black solid curve stands for the unitary dynamics $P_{rr}(t)$ in the case of effective system as same as presented in Fig.\ref{exam}(a).  }
\label{oudys}
\end{figure}

By then we can calculate the original time-dependent unitary dynamics between the ground state $|11\rangle$ and the doubly-excited state $| rr\rangle$ in an exact numerical way, as shown in Fig. \ref{oudys} where the results from the effective system are plotted by black solid curves for contrast. Clearly when the coupling field $\Omega_s$ is modulated timely, a close consistence between the unitary dynamics of two models (original and effective) can be preserved, enabling a sufficient residence time on $| rr\rangle$ for the dissipation of population onto the target state. That fact could be understood by the well definition of parameters $\Omega(t) = \Omega_s(t)\Omega_p/2\delta= (\Omega_{p0}\Omega_{s0}/2\delta)\cos(\omega_0t)$, $\Delta(t) = (\Omega_{p0}^2 - \Omega_{s0}^2\cos^2(\omega_0t))/4\delta\approx\Omega_{p0}^2/4\delta$=const due to $\Omega_{p0}\gg\Omega_{s0}$. However if the pump laser $\Omega_p(t)$ is modulated, although $\Omega(t)$ is also well-defined yet $\Delta(t) =(\Omega_{p0}^2\cos^2(\omega_0t) - \Omega_{s0}^2)/4\delta\approx(\Omega_{p0}^2/4\delta)\cos^2(\omega_0t) $ becomes a strongly oscillating function with amplitude $\Omega_{p0}^2/4\delta$ that can not keep the requirement of a big detuning to the singly-excited states as time evolves, arising a significant destruction for the unitary dynamics of $|11\rangle\leftrightarrows |rr\rangle$, since the effective system by discarding all singly-excited states is unreasonable then. Therefore we can robustly verify the validity of our simpler effective scheme compared with a real system in the experimental implementation.

\end{document}